\documentclass[a4paper,11pt]{article}
\usepackage{jheppub} % for details on the use of the package, please see the JINST-author-manual
\usepackage[english]{babel}
\usepackage{multirow}
\usepackage{multicol}
\usepackage{array}
\usepackage{soul}
\usepackage{amsmath}
\usepackage{ulem}
\usepackage{graphicx}
\usepackage{caption}
\usepackage{subcaption}
\usepackage{graphicx}
\usepackage{verbatim}
\title{Search for Stochastic GW Signal as a Complementary Approach to Multi-Higgs Productions at the Hadron Colliders to Probe Dimension Six Operator}
\author[a]{Dilruba Gazi,}
\author[b]{Abhi Mukherjee,} 
\author[b]{Saurabh Niyogi,}
\author[a]{Sujoy Poddar}
\affiliation[a]{Department of Physics, Diamond Harbour Women’s University, Diamond Harbour Road, Sarisha, South 24 Parganas, West Bengal 743368, India}{}
\affiliation[b]{Department of Physics, Gokhale Memorial Girls' College, 1/1 Harish Mukherjee Road, Kolkata 700 020, India}{}

\emailAdd{dilruba.gazii@gmail.com}
\emailAdd{abhi.mukherjee006@gmail.com}
\emailAdd{saurabhphys@gmail.com}
\emailAdd{sujoy.phy@gmail.com}

\abstract{ 
We have considered an effective field theory framework in which the Standard Model is extended by a non-renormalizable dimension six operator ($\frac{1}{\Lambda^2}(H^{\dagger}H)^3$) respecting the symmetries of the Standard Model. Such an operator can affect the dynamics of Higgs field and the electroweak phase transition. Presence of such term modifies the triple Higgs coupling which is already under the lens from di-Higgs searches in various channels at the LHC. Constraint on the scale parameter $\Lambda$ is obtained from the di-Higgs data at the LHC Run II. Additionally, such dimension six terms will give rise to additional couplings which will affect tri-Higgs and four-Higgs productions at future high energy, high luminosity hadron colliders. The shape of the scalar potential also gets modified due to such higher dimensional term composed of Higgs field. It is worthwhile to study the impact of such higher-dimensional operators at suitable gravitational wave (GW) interferometry experiments since GW astronomy has become a good probe for new physics searches.
Bounds obtained on the scale parameter $\Lambda$ from the allowed first-order electroweak phase transition is $450~\text{GeV} \lesssim \Lambda \lesssim 590$~GeV. On the other hand, di-Higgs results from the LHC  Run II put a lower bound on $\Lambda \gtrsim 340$ GeV.
}

\begin{document}
\maketitle

\section{Introduction}
%%%
The spin zero positive parity ($0^+$) particle discovered at the Large Hadron Collider (LHC) \cite{ATLAS:2012yve,CMS:2012qbp} is increasingly seeming to be `the' Higgs boson conjectured in the Standard Model (SM), as almost all the properties concerning the particle found from the LHC searches for over a decade match with the SM predictions. With the chance of discovering new physics (NP) in the form of new resonances gone dry, one may start to believe to be looking at the particle desert. In such a scenario, the effective field theory (EFT) provides a very useful guidance. In the context of the Standard Model Effective Field Theory (SMEFT), higher dimensional operators are those that extend the SM by introducing new terms in the Lagrangian respecting its Lorentz and Gauge symmetries, but are suppressed by a scale ($\Lambda$), typically associated with the scale of new physics beyond the SM. These operators have mass dimension greater than four and are constructed out of the SM fields \cite{Giudice:2007fh,Grzadkowski:2003sd,Bodeker:2004ws,Grzadkowski:2010es,Jenkins:2013zja,Contino:2013kra,Alonso:2013hga}. The general form of such operators is:
\begin{equation}
 \mathcal{O}_i = \frac{c_i}{\Lambda^{n-4}} \mathcal{O}^{(n)}_{\text{SM}, i},   \nonumber
\end{equation}
where \(c_i\) are dimensionless coefficients, \(\Lambda\) is the scale of NP, and \(\mathcal{O}^{(n)}_{\text{SM}, i}\) are the higher-dimensional operators constructed from the SM fields.

In this work, we consider an operator having dimension six constructed solely from the Higgs field. Among a number of dimension six operators, \(\mathcal{O}_H = \frac{1}{\Lambda^2}(H^{\dagger}H)^3\) is a natural choice (here, $H$ stands for Higgs field) because it has specific effects on various observables related to the Higgs sector of the SM, particularly, the Higgs potential and can modify the Higgs self-coupling, which is a topic of great interest at the LHC. In this paper, henceforth, we shall mention this scenario as SM+H6 model. Apart from the self-coupling, which is already under the lens at the LHC, the shape of the Higgs potential plays a crucial role in some early universe cosmological phenomena, like cosmic inflation and the electroweak phase transition (EWPT) \cite{Dine:1992wr, Kajantie:1995kf, Kajantie:1995dw, Moore:2000jw}. In the SM, the transition is a smooth cross over, meaning it does not generate significant gravitational waves (GW). However, the dimension-6 operator can modify the Higgs potential such that it induces first-order phase transition. In this work, we are particularly interested in studying the effect of such modifications in the EWPT in the early universe and the subsequent emergence of the GWs which could be observed in a number of space as well as ground based interferometry experiments. 

It is understood that ignoring other dimension six operators may lead to an incomplete understanding of the NP effects. Many other operators can affect Higgs phenomenology \cite{Anisha:2021fzf}, electroweak precision observables \cite{DasBakshi:2020pbf}. However, by focusing on a particular operator our goal is to aim at isolating specific effects and gaining insights into the potential impact of NP on the Higgs phenomenology. Needless to say, this obviates many theoretical complexities and simplifies interpretations in terms of the results since including only one operator (or, a few operators) at a time reduces the number of free parameters in the effective theory, making it easier to fit the experimental data and to draw specific conclusions.

%\sout{Detection of GWs has opened up the possibility of search strategies for various particle physics models either alternative or complementary to the collider searches. Among many such possible GW sources in the early universe, cosmic phase transitions \cite{Witten:1984rs,Hogan:1986dsh} have been extensively studied. }%It is also well-known that EWPT in the early universe is one of the three Sakharov conditions \cite{} for generating the observed baryon asymmetry in the universe.

With no new particle in sight at the LHC, it is logical to take effective theory approach which encapsulates the effects of the unknown new physics in the form of some higher dimensional operators which are otherwise absent in the SM due to the pre-condition of renormalizability of the theory. With the goal in mind to study the effects of the EWPT in the early universe which may generate a detectable stochastic GW in addition to effects of Higgs sector physics already undergoing at the LHC simultaneously under the EFT framework, the natural first choice is the dimension six $(H^{\dagger}H)^3$ term augmenting the SM scalar potential. Besides, such a term can also be automatically generated, for instance, from some new strongly interacting sector at some high scale, or by integrating out the heavy particles. The inclusion of this operator modifies the Higgs self-coupling, which in turn, gets constrained from di-Higgs searches at the LHC. In this work, we try to assemble all such pieces together and try to distill useful information. However, the most promising part of this work is the prediction of tri-Higgs production from gluon fusion and search \cite{Plehn:2005nk,Binoth:2006ym,Brigljevic:2024vuv} in presence of $(H^{\dagger}H)^3$ term at the LHC and Future Circular Hadron Collider (FCC-hh) machine \cite{Contino:2016spe,FCC:2018byv,Fuks:2015hna}. More recently, study of multiple Higgses in the collider has been undertaken in \cite{Brigljevic:2024vuv,Biermann:2024oyy}. However, multiHiggs production does not seem a viable possibility at the LHC, rather it needs higher center-of-mass energy and higher luminosity collider, which is not immediately in sight. 

Meanwhile, the possibility of GW detection in various interferometer based experiments could throw up a nice alternative to test such new physics effect hidden under the dimension six operator under consideration.
The effect on the GW spectrum due to $(H^{\dagger}H)^3$ term and its possible observation in various GW detectors like, LISA \cite{Caprini:2015zlo, Athron:2023xlk}, BBO \cite{Crowder:2005nr, Corbin:2005ny, Harry:2006fi}, DECIGO \cite{Seto:2001qf, Kawamura:2006up,Isoyama:2018rjb} have been studied earlier \cite{Grojean:2004xa,Delaunay:2007wb,Ellis:2018mja,Banerjee:2024qiu} \footnote{During the final stage this work, an article \cite{Qin:2024dfp} with partially similar analysis appeared in the arXiv.}.  
 Therefore, time is ripe for new search strategies where signatures from early universe provide substitute for searches in colliders, which are yet to come \cite{Profumo:2007wc,Profumo:2014opa,Huang:2017jws,Chen:2017qcz,Carena:2018vpt,Alves:2019igs,Ghosh:2022fzp,Alves:2020bpi,Roy:2022gop}. In addition, one must realize that presence of $(H^{\dagger}H)^3$ term in the Lagrangian would, in principle, allows a vertex involving maximum of six Higgses which would contribute to the production of five Higgses in addition to the SM contribution. In this work, we have considered four-Higgs production comprising the five Higgs vertex exclusively arising from SM+H6 scenario.

The paper is organised as follows. In sec.\ref{sec-2}, we briefly discuss the one-loop effective potential for the dimension six operator. The constraints from di-Higgs searches at the LHC have been discussed in sec.\ref{sec-3}. A brief thermal history of the universe and the generation of GW spectra from the collisions of bubbles of newly formed electroweak phase are discussed in sec.\ref{sec-4}. The next sec.\ref{sec-5} consists of tri-Higgs and four-Higgs searches at the LHC and FCC-hh colliders. Results from GW astronomy are discussed in sec.\ref{sec-6}. Finally, we summarize in sec.\ref{sec-7}.

\section{One-Loop Effective Potential with a Dimension Six Operator $(H^{\dagger}H)^3$}\label{sec-2}

The effective potential is a crucial tool for studying the vacuum structure of a theory, particularly in the context of spontaneous symmetry breaking. The effective potential V$_{eff}$ includes quantum corrections and is essential for understanding the behavior of fields beyond the classical approximation.

\subsection{Tree level Potential}
The classical scalar potential V$_0$(H) for the Higgs field $H$ with the additional dimension six term is given by:
\begin{equation}\label{eqn:scalar_pot}
V_0(H) = -\mu^2~ (H^{\dagger}H) + \lambda~(H^{\dagger}H)^2 + \frac{1}{\Lambda^2}~(H^{\dagger}H)^3,
\end{equation}
where $\mu$ is the bare mass parameter, $\lambda$ is the quartic coupling constant and $\Lambda$ is the scale of new physics, signifying the validity of the theory. From the potential given in Eq.\ref{eqn:scalar_pot}, it is obvious that $\mu, \Lambda$ each has mass dimension one, whereas, the quartic coupling $\lambda$ is a dimensionless quantity. Considering 
$H = \frac{1}{\sqrt{2}}\begin{pmatrix}
0 \\
h + v
\end{pmatrix}$,
one may expand the terms in the potential \ref{eqn:scalar_pot} and estimate required couplings in the Higgs sector.  From the three parameters in eqn.\ref{eqn:scalar_pot}, we trade in $\mu^2$ and $\lambda$ using the minimization condition and mass determining condition, namely, $V^{'}_{0} (h = v) = 0$ and $V^{''}_{0} (h = v) = m^2_h$ to obtain the following expressions \cite{Delaunay:2007wb,Ellis:2018mja}:
\begin{equation}\nonumber
\mu^2 = \frac{m^2_h}{2} - \frac{3v^4}{4\Lambda^2},~~~ \lambda = \frac{m^2_h}{2v^2} - \frac{3v^2}{2\Lambda^2}
\end{equation}
Therefore, we fix $m_h = 125$ GeV and $v = 246$ GeV, only to consider $\Lambda$ as the free parameter in the theory.
%%%%

\subsection{One loop effective potential}
The one-loop effective potential $V_{eff}$ includes quantum corrections to the classical potential. Three contributions in the effective potential are the following \cite{Quiros:1999jp}:

\begin{equation}\label{eq:veff}
V_{eff} (H) = V_0(H) + V_1^{T=0}(H) + V_1^{T\ne0}(H)
\end{equation}
where $V_1^{T=0}(H)$ is the one loop correction term at $T=0$ temperature which is given by the Coleman-Weinberg formula \cite{PhysRevD.7.1888}:
\begin{equation}
V_1^{T=0}(H) = \pm \frac{1}{64 \pi^2} \sum_{i = h,W,Z,t} ~ n_i M^4_i~\left( \text{ln}~\frac{M^2_i(H)}{M^2_0} - c_i \right)
\end{equation}
with 
    $M_i$ being the field-dependent masses,
    $n_i$ representing the degrees of freedom,
    $M_0$  is the renormalization scale,
    $c_i$ are numerical factors ($c_i = \frac{3}{2}$ for fermions and $\frac{5}{6}$ for bosons in the $\overline{MS}$ scheme),
    $+$ve ($-$ve ) sign is for bosonic (fermionic) degrees of freedom.

The finite temperature part $V_1^{T\ne0}(H)$ is given by
\begin{eqnarray} \nonumber
V_1^{T\ne0}(H) &=& \sum_i~ \frac{n_i T^4}{2 \pi^2}~\int^{\infty}_0 ~k^2~\text{ln} \left( 1 \mp 
e^{-\sqrt{k^2 + m^2_i(H)/T^2}} \right)~dk \\
&\equiv& \sum_{i=\text{bosons}}~\frac{n_i T^4}{2 \pi^2} J_b\left( \frac{m^2_i(H)}{T^2}\right) + \sum_{i=\text{fermions}}~\frac{n_i T^4}{2 \pi^2} J_f\left( \frac{m^2_i(H)}{T^2}\right)
\end{eqnarray}
where $J_{b/f}$ are the integral
\begin{equation}
J_{b/f}\left( \frac{m^2_i}{T^2} \right)  = \pm \int^{\infty}_0~ dy~y^2~\text{ln}\left( 1 \mp e^{-\sqrt{y^2 + \frac{m^2_i}{T^2}}} \right), \nonumber
\end{equation}
where the upper (lower) sign is for bosons (fermions). $n_i$s are the degrees of freedom corresponding to various fields whose numerical values are $n_{h, W, Z, t} = 1, 6, 3, 12$ respectively.  
%\subsection{ \textcolor{red}{Field Dependent Masse}}
\subsection{Contribution from Ring Diagrams}
%The masses $M_i$s are functions of the field $H$. Considering $H = \frac{1}{\sqrt{2}}\begin{pmatrix}
%0 \\
%h + v 
%\end{pmatrix}$ 

To enhance the accuracy of our analysis, we must incorporate a correction arising from the resummation of multi-loop contributions involving the longitudinal polarizations of bosons \cite{Parwani:1991gq,Carrington:1991hz,Arnold:1992rz,Arnold:1992fb}. This adjustment necessitates shifting the masses of these longitudinal polarizations of both gauge bosons and scalars by their corresponding thermal corrections. Consequently, the original mass parameter $m^2_i$ is modified to ($m^2_i + \Pi_i$).

Within the framework of this model incorporating of dimension six operator, these shifts are given by \cite{Carrington:1991hz,Delaunay:2007wb}:
\begin{eqnarray}
    \Pi_{h,\chi_i}(T) = \frac{T^2}{4v^2}(m_h^2+2m_W^2+m_Z^2+2m_t^2) - \frac{3}{4} T^2 \frac{v^2}{\Lambda^2}
\end{eqnarray}
\begin{eqnarray}
    \Pi_W(T) = \frac{22}{3} \frac{m_W^2}{v^2}T^2
\end{eqnarray}

with the shifted masses of $Z$ and $\gamma$ are the eigenvalues of the following mass matrix \\

\begin{center}
$\begin{pmatrix}
    \frac{1}{4}g^2 h^2+\frac{11}{6}g^2T^2 &  - \frac{1}{4} g^{\prime 2} g^2 h^2 \\
    - \frac{1}{4} g^{\prime 2} g^2 h^2 &  \frac{1}{4}g^{\prime 2} h^2+\frac{11}{6}g^{\prime 2}T^2 
\end{pmatrix}$
\end{center}
The fields $h, \chi, W, Z, \gamma$ are the physical Higgs field, the Goldstone bosons, weak bosons and photon respectively and $m_h$, $m_W$, $m_Z$  and $m_t$ are the masses of physical Higgs, $W$ boson, $Z$ boson and top quark respectively. Further, $g$ and $g^{\prime}$ are the SM $SU(2)$ and $U(1)$ gauge couplings respectively.
These modifications are essential to ensure the reliability and accuracy of our theoretical predictions, particularly in high-temperature environments or when considering quantum field theory effects at finite temperatures.

Finally, the complete temperature-dependent renormalized effective potential at one-loop is given by:

\begin{eqnarray}\label{eq:veff_total}
    V_{eff}  = & - &\mu^2~ h^2 + \lambda~ h^4 + \frac{1}{\Lambda^2}~ h^6 \pm  \frac{1}{64 \pi^2} \sum_{i = h,\chi,W,Z,t} ~ n_i M^4_i~\left( \text{ln}~\frac{M^2_i(h)}{M^2_0} - c_i \right) \nonumber \\  & + & \sum_{i=\text{bosons}} ~  \frac{n_i T^4}{2 \pi^2} J_b\left( \frac{m^2_i(h)}{T^2}\right) + \sum_{i=\text{fermions}}~\frac{n_i T^4}{2 \pi^2} J_f\left( \frac{m^2_i(h)}{T^2}\right) \nonumber \\ & + & \sum_{i = h,\chi,W,Z,\gamma} ~ \frac{\bar{n}_i T}{12 \pi} ~\left[ m_i^3(h) - \left( m_i^2(h) + \Pi_i(T)\right)^{3/2} \right]
\end{eqnarray}
\noindent
where, $\bar{n}_{(h,\chi,W,Z,\gamma)} = (1,3,2,1,1)$. This is the potential we have analyzed to explore first-order phase transitions and gravitational waves within the context of the dimension six extension of the SM for the remainder of this article. We have used {\tt CosmoTransitions} \cite{Wainwright:2011kj} which is a numerical package to analyze phase transitions using the finite temperature field theory with single or multiple scalar fields.

%%%%%%%%%%%%%%%%%%%%%%%%%%%%%%%%%%%%%%%%%%%%%%%%%%%%%%%%%%%%%%%%%%%%5
\begin{figure}[htb]
    \centering
    \includegraphics[width=0.45\linewidth]{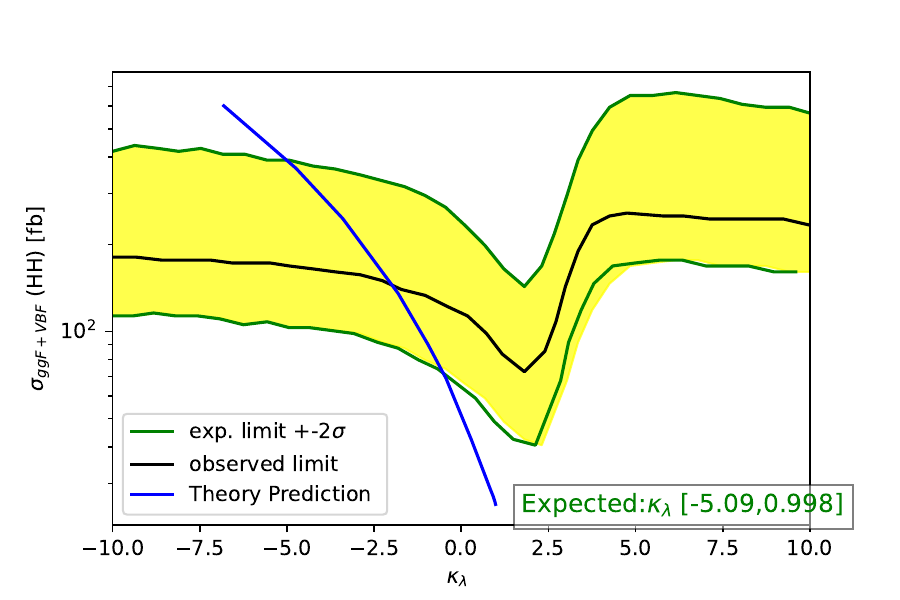}
     \includegraphics[width=0.45\linewidth]{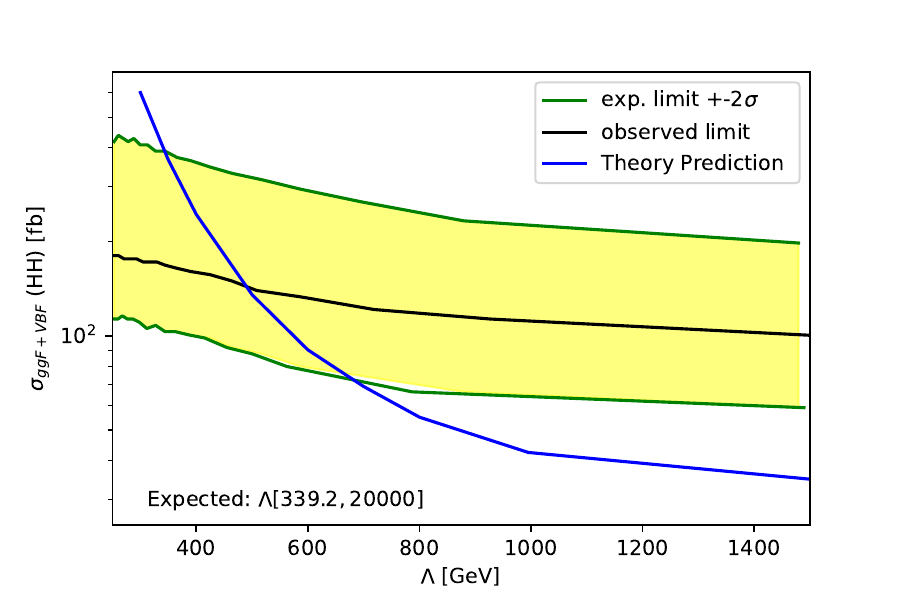}
    \caption{   
    The central black line in left panel shows the di-Higgs production cross-section measured by the ATLAS collaboration \cite{ATLAS:2021tyg} as a function of Higgs self-coupling modifier $\kappa_{\lambda} = \frac{ \lambda^{SM+H6}_{hhh}}{\lambda^{SM}_{hhh}}$. The yellow band denotes $\pm 2\sigma$ uncertainty around the observed value. The blue curve displays the di-higgs cross-section in SM+H6 scenario. The figure in the right panel shows the allowed region where the theoretical blue curve falls below the experimental $2\sigma$ band. It can be seen that the line goes asymptotically
    to arbitrary large $\Lambda$ which corresponds to the SM.}
    \label{fig:lhc_dihiggs}
\end{figure}

\section{Constraints from di-Higgs searches at the LHC Run II}\label{sec-3}
%%%%%%%%%%
The interferometer-based gravitational wave astronomy experiments is a way to ``observe" the stochastic GW relics created during the early universe phase transition. In a very similar manner, one may think of the LHC as a man-made machine to recreate an early universe environment on the earth. Therefore, to measure the properties of the scalar particle discovered at the LHC is of paramount importance in the post-discovery era of the SM-like Higgs boson as it unveils the possible structure of the scalar potential \cite{Agrawal:2019bpm}. The Higgs couplings to the SM electroweak gauge bosons are in excellent agreement with the SM expectations, as already confirmed by the LHC Run I and Run II data \cite{ATLAS:2023tnc,ATLAS:2023pwa, CMS:2022uhn, CMS:2023gjz}. The rarity of events due to extreme smallness of the Higgs couplings to the first two generations of quarks and leptons, the discovery of those couplings can not be established yet at the LHC \cite{ATLAS:2022ers, CMS:2022fxs}. However, a circumspect progress has already been made in the couplings to the third generation quarks and lepton \cite{ATLAS:2024wfv,ATLAS:2024moy,ATLAS:2021qou,CMS:2022dbt,CMS:2024fdo}. The only other measurable coupling which is also of utmost importance in describing the shape of the scalar potential of the theory is the Higgs self-coupling ($\lambda_{hhh}$). Within the framework of the SM, the value of this coupling is not small at all, however, this self-coupling can be measured only from events where a single Higgs Boson decays into two Higgs Bosons. Additionally, the two Feynman diagrams for the di-Higgs process, the box diagram and the triangle diagram, interfere destructively, resulting in low cross-section for the process. The non-resonant Higgs pair production cross-section for a Higgs mass of $m_h = 125$ GeV for gluon-gluon-fusion (ggF) process in the SM at next-to-next-to-leading order (NNLO) is $\sigma^{\text{SM}}_{\text{ggF}} = 31.1$ fb at the center-of-mass energy of $13$ TeV at the LHC \cite{Dawson:1998py,Shao:2013bz,deFlorian:2013jea,deFlorian:2015moa,Borowka:2016ehy,Grazzini:2018bsd,Baglio:2018lrj,Baglio:2020wgt}. The same for vector-boson-fusion (VBF) process is $\sigma^{\text{SM}}_{\text{VBF}} = 1.73 \pm 0.04$ fb at N$^3$LO in QCD \cite{Baglio:2012np,Frederix:2014hta,Ling:2014sne,Dreyer:2018qbw,Dreyer:2018rfu}. In this article, we impose the constraint considering the results from the ATLAS di-Higgs searches \cite{ATLAS:2024pov,ATLAS:2024lsk,ATLAS:2024ish}.

 The cross-sections have been displayed in Fig.\ref{fig:lhc_dihiggs} where the plot in the left panel shows the latest bound on self-coupling modifier $\kappa_{\lambda}$, defined as $\kappa_{\lambda} =\frac{ \lambda^{SM+H6}_{hhh}}{\lambda^{SM}_{hhh}}$. The values of $\kappa_{\lambda}$ above the $2\sigma$ region of experimental contour is disallowed. However, the region below yellow band is still allowed. The expected allowed region for the self-coupling modifier is $-5.09 < \kappa_{\lambda} \lesssim 0.998$, where the extreme right limit corresponds to $\Lambda \sim 20$ TeV, which reaches SM value asymptotically. The right panel of Fig. \ref{fig:lhc_dihiggs} shows the lower bound on the energy scale as $\Lambda \gtrsim 340$ GeV for the dimension six operator in Eq.\ref{eqn:scalar_pot}. This is obtained using the results of the searches performed at LHC Run II using di-Higgs data set for $(126-139)$ fb$^{-1}$ of integrated luminosity.

\begin{figure}[htb]
    \centering
    \includegraphics[width=0.65\linewidth]{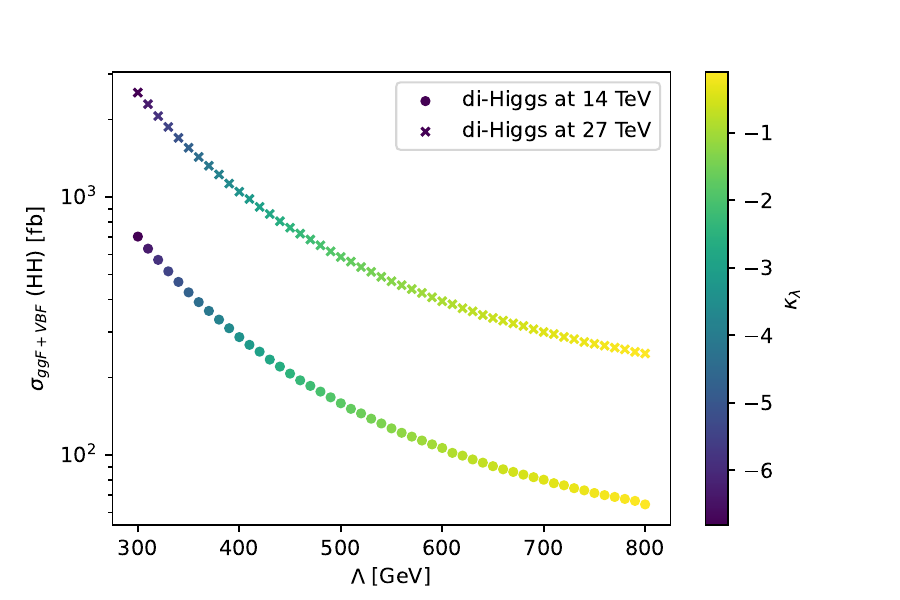}
    \caption{Cross-section for di-Higgs production considering ggF and VBF with a renormalization scale $\mu_0=\frac{M_{hh}}{2}$, where $M_{hh}$ is the di-Higgs invariant mass with $m_h=125 $ GeV at center-of-mass energies $14$ TeV at the LHC and $27$ TeV HE-LHC against the new physics scale $\Lambda$ in SM+H6 scenario. Both ggF as well as VBF processes are included in the estimation of the di-Higgs cross-section. The color gradient on the right side of the panel presents the values of ${\kappa}_{\lambda}$.}
    \label{fig:lhc_plot2}
\end{figure}

The projected di-Higgs cross-sections at $14$ TeV LHC for SM+H6 scenario are shown in Fig.\ref{fig:lhc_plot2}. The di-Higgs cross-section has been estimated from the combined numbers of the gluon-gluon fusion and vector boson fusion processes. Furthermore, we have also shown the di-Higgs cross-section at $27$ TeV at the HE-LHC in Fig.\ref{fig:lhc_plot2}. 
Towards this we implement the scenario in {\tt MadGraphaMCatNLO} \cite{Alwall:2011uj} using {\tt FeynRules v2.3} \cite{Christensen:2008py} in conjunction with NLOCT \cite{Degrande:2014vpa} for incorporating loop diagrams via its {\tt UFO} (Univeral Feynrules Output) \cite{Degrande:2011ua,deAquino:2011ub} interface. 
%Since the lowest order Feynman diagrams of all the processes involve loop diagrams, we employ {\tt NLOCT} \cite{Degrande:2014vpa} inside the {\tt Feynrules}. 
The parton distribution distribution (PDF), in this case, is considered to be NN23LO \cite{Ball:2013hta, NNPDF:2014otw}.
%%%%%%%%%%%%%%%%%%%%%%%%%%%%%

\section{Thermal History of the Universe: SFOPT and GW}\label{sec-4}

The idea of EWPT begins following the Big Bang theory that the universe in the remote past is believed to be at much higher temperature at which the SM electroweak gauge symmetry remains unbroken, and the Higgs VEV is zero. As the temperature drops due to the expansion of the universe, the structure of the scalar potential changes in such a way that at some point, the VEV becomes non-zero, thereby breaking the $SU(2) \times U(1)_Y$ gauge symmetry. The transition from $\langle H \rangle = 0$ to $\langle H \rangle \ne 0$ proceeds via spontaneous nucleation of bubbles of the new phase. The bubbles expand at the expense of free energy i.e. the energy difference between the two minima. The expanding bubbles collide percolating the new phase throughout the entire Hubble volume, finally releasing the latent heat of the phase transition, reheating the universe in the process and marking the end of transition. Such collisions of many bubbles, violating the isotropy of the universe, result in the generation of stochastic GWs. In first order phase transitions, the sources of GWs are classified into the following \cite{Caprini:2015zlo}: bubble wall collisions \cite{Kosowsky:1991ua,Kosowsky:1992rz,Kosowsky:1992vn,Kamionkowski:1993fg,Caprini:2007xq,Huber:2008hg}, turbulence \cite{Dolgov:2002ra,Nicolis:2003tg,Caprini:2006jb,Gogoberidze:2007an,Kahniashvili:2008er,Kahniashvili:2008pe,Megevand:2008mg,Kahniashvili:2009mf,Caprini:2009yp} and sound waves \cite{Hindmarsh:2013xza,Kalaydzhyan:2014wca,Hindmarsh:2015qta}. The first occurs due to the scalar field configuration, while the last two originate from the dynamics in the thermal plasma inside the unbroken phase.

%\begin{figure}[htb]
%	\centering
%\subfloat[]	{\includegraphics[width=0.49\textwidth]{lambda490.pdf}}\label{fig:A1}
%\subfloat[]	{\includegraphics[width=0.49\textwidth]{lambda590.pdf}}\label{fig:B1} 
%\caption{\textcolor{red}{What do we say about this plots?? Do we understand that FOPT is not possible beyond $\Lambda = 490, \Lambda = 590$ GeV from these %two graphs??}}
%	\label{fig:vphi} 
%\end{figure}

Spontaneous nucleation of bubbles of new phase ($V = V_{\rm true}$) or equivalently, the decay false vacuum to the true one, starts due to thermal fluctuations at finite temperature \cite{Linde:1977mm, Linde:1978px} and due to quantum tunneling at zero temperature \cite{Coleman:1977py, Callan:1977pt}. At the onset of the true vacuum, the scalar field still remains trapped in the false vacuum due to the insufficient energy $\Delta V = V_{\rm false} - V_{\rm true}$, which drives the thermal transition \cite{Espinosa:2010hh}. In the context of cosmology, the bubble starts to nucleate when probability of transition (i.e. bubble nucleation) per unit Hubble volume per unit Hubble time becomes $\sim \mathcal{O}(1)$. In specific models, such transition can be delayed and universe remain in the false vacuum for some time, even though the newly appeared minimum (the true vacuum) is thermodynamically favorable. Such a supercooled state of the universe is desirable in order to generate high frequency GW signal with from bubble collision. However, with the universe ever expanding, a prolonged supercooled state would not allow to complete the transition throughout the universe.

For the EWPT to be a SFOPT, the Higgs potential at finite temperature must have a significant barrier between the symmetric phase (where the Higgs field VEV is zero) and the broken phase (where the Higgs field VEV is non-zero). This barrier allows for the nucleation of bubbles of the broken phase within the symmetric phase, which is critical for a strong first-order phase transition. Strong first-order phase transition to occur in the framework of SM, the Higgs boson mass would need to be below about $\sim 70$ GeV \cite{Bochkarev:1987wf,Morrissey:2012db}, which is well below the observed mass. However, the dimension six operator $(H^{\dagger}H)^3$ augmenting the scalar potential helps in creating the barrier with the observed Higgs mass at $125$ GeV. Note that only a small range of $\Lambda$ is favourable for EWPT as can be observed from tab.\ref{tab:my_label}. For $ 450~\text{GeV} < \Lambda < 590 $ GeV, EWPT proceeds via successful nucleation of bubbles of broken  phase of electroweak symmetry. $\Lambda$ beyond these ranges are not conducive for EWPT.

\begin{table}[]
    \centering
    \resizebox{0.5\textwidth}{!}{
    \begin{tabular}{||c|c|c|c||}
    \hline
    \hline
      $\Lambda$ (GeV)   &  $\alpha$ & $\beta/H_n$ & $T_n$ (GeV) \\
      %\multicolumn{4}{||c||}{No transition found} \\
      \hline
       590 & 0.2909 & 215.38 & 39.14 \\
       
       550 & 0.0528 & 1608.20 & 97.32 \\
       
       500 & 0.0232 & 3922.72 & 147.10 \\

       450 & 0.0069 & 9738.97 & 204.24 \\
       \hline
        \hline
        %\multicolumn{4}{||c||}{No transition found} \\
    \end{tabular} }
    \caption{Variations of two dimensionless parameters $\alpha$ and $\beta/H_n$, quantifying the quality of GW spectra (defined in sec. \ref{sec-6}), and the nucleation temperature $T_n$ against the energy scale $\Lambda$. For $\Lambda \lesssim 450$ GeV and $\Lambda \gtrsim 590$ GeV, bubble nucleation does not take place which means EWPT does not occur beyond these values of $\Lambda$s.}
    \label{tab:my_label}
\end{table}

\section{Multi-Higgs Productions at the HE-LHC and FCC-hh}\label{sec-5}
%%%%%
%%
\subsection{Tri-Higgs Production}

\begin{figure}[h!]
	\centering
\subfloat[]	{\includegraphics[width=0.40\textwidth]{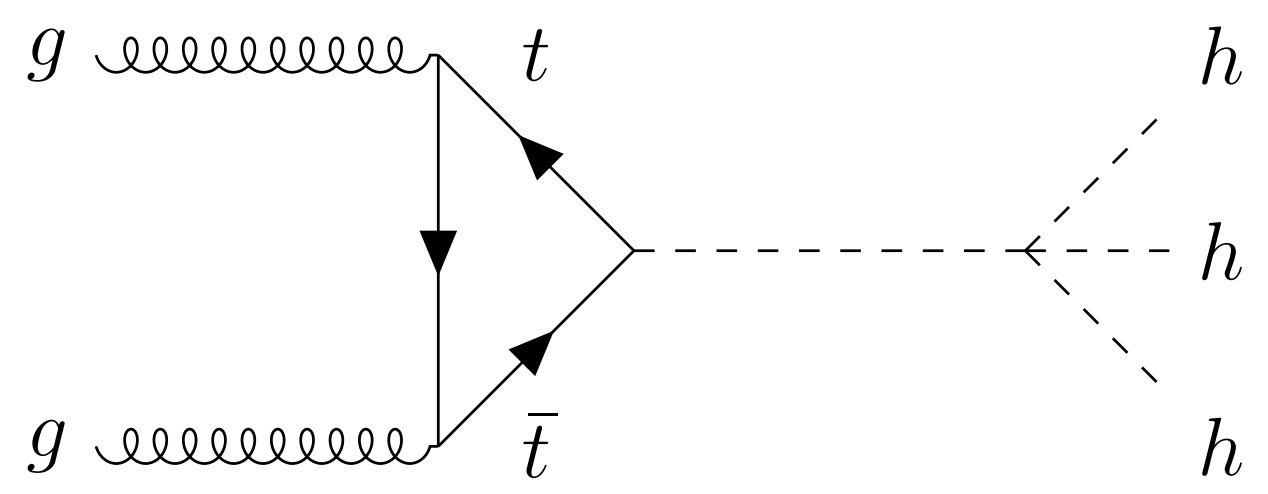}}
\hspace{1cm} \subfloat[]	{\includegraphics[width=0.40\textwidth]{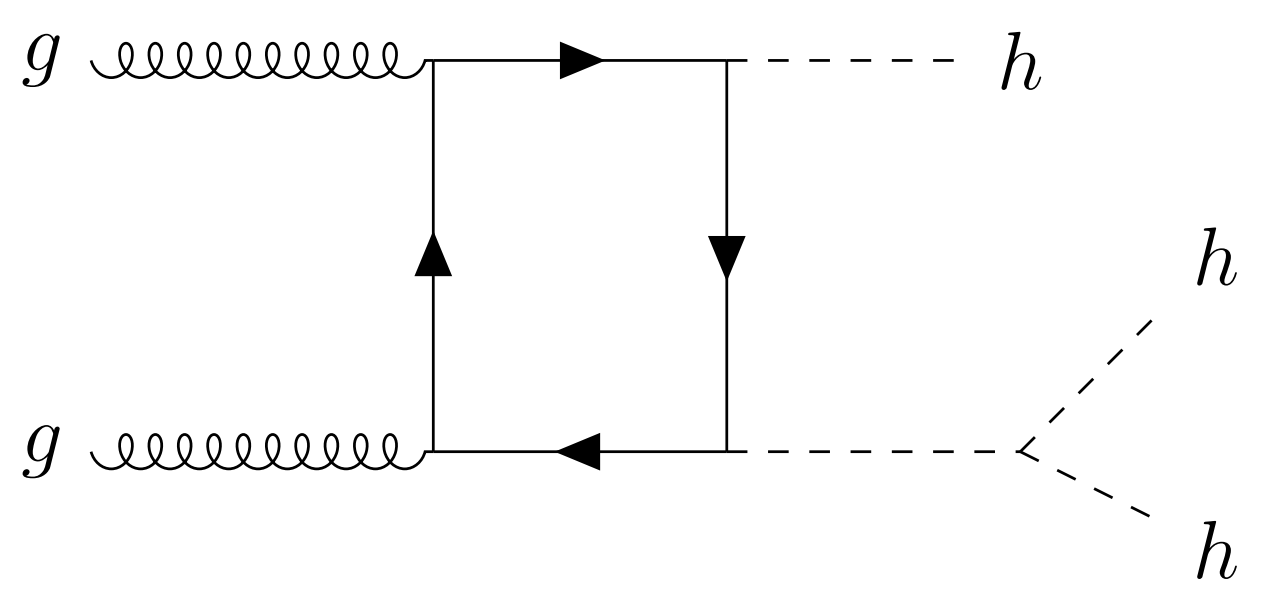}} \\
\subfloat[] {\includegraphics[width=0.40\textwidth]{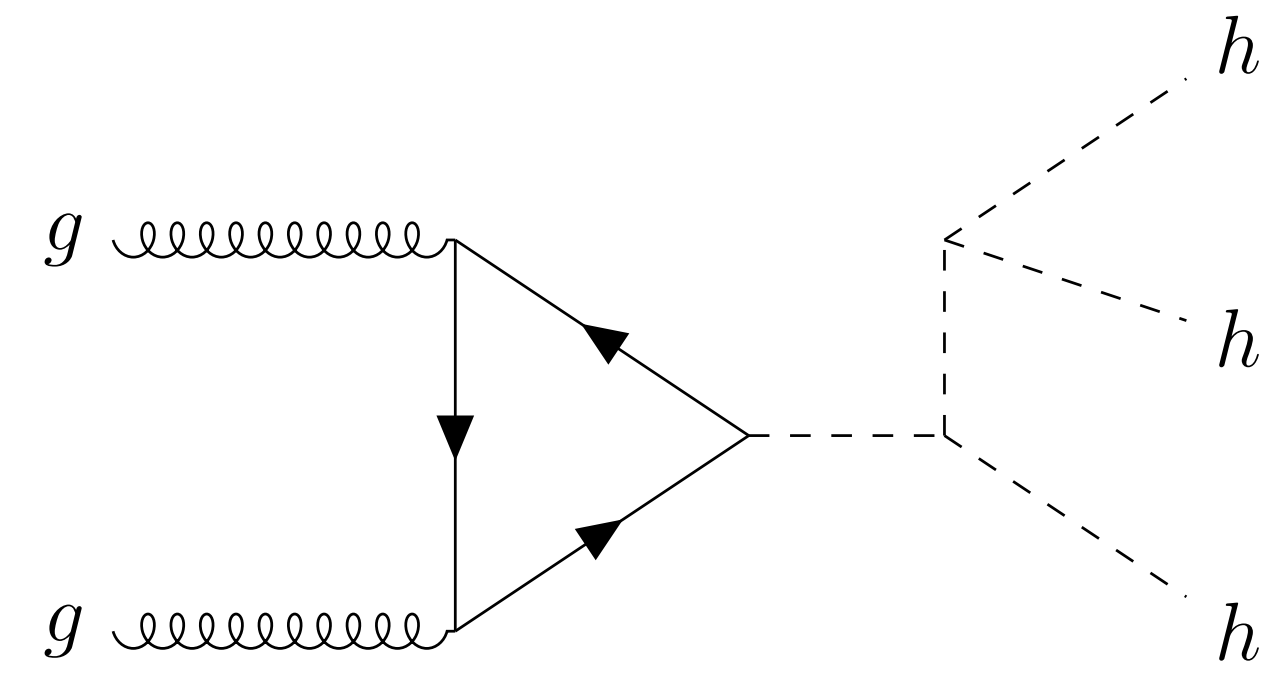}}
\caption{Representative Feynman diagrams contributing to the cross-section for the process $gg \rightarrow hhh$. Only top and bottom quarks have been considered inside the loops.} 
	\label{fig:hhhfeynman} 
\end{figure}

With theoretical and experimental results available aplenty, Higgs pair production at the LHC is now an established research topic. On the other hand, the study of $hhh$ production is still a rather new field of research. %largely uncharted territory.%
While some theoretical estimations do exist \cite{Plehn:2005nk, Binoth:2006ym, Maltoni:2014eza, Fuks:2017zkg, Papaefstathiou:2019ofh}, attention from the experimental community is lacking due to small cross-section at the LHC. The reason is, of course, very little cross-section even at the highest center-mass-energy energy available so far.  The production of triple Higgs receives contributions from similar triangle and box diagrams of di-Higgs production, with an additional vertex where one of the final state Higgses goes into two more Higgses shown in Fig.\ref{fig:hhhfeynman}. Such contributions are dominant. 
%%%%%%
%%%%%

\begin{figure}[htb]
        \centering
        \includegraphics[width=0.7\linewidth]{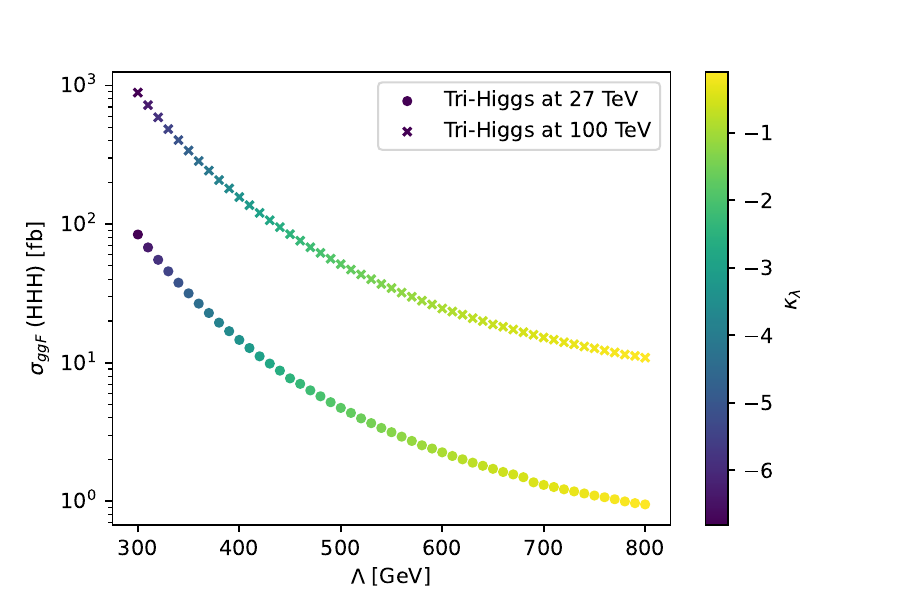}
      \caption{{Tri-Higgs production cross-sections via ggF ($gg \rightarrow hhh$) is plotted against the scale parameter $\Lambda$ for center-of-mass energies $27$ TeV HE-LHC and $100$ TeV FCC-hh. Different symbols as shown in the figure are used to differentiate the two different colliders. The renormalization scale $\frac{M_{hhh}}{2}$, where $M_{hhh}$ is the tri-Higgs invariant mass with $m_h=125 $ GeV}, is used for computing the cross-sections.}
      %Tri-Higgs cross section at $27$ TeV and $100$ TeV colliders} \label{fig:trihiggs-xsec}
   \label{fig:trihiggs}
\end{figure}

Furthermore, quartic Higgs coupling ($\lambda_{hhhh}$) in SM provides another Feynman diagram to furnish some additional contributions. However, in SM+H6 model, both the $\lambda_{hhh}$ and $\lambda_{hhhh}$ couplings get altered with respect their SM values due to presence of $H^6$ term. It can be empirically understood from the magnitudes of triple Higgs cross-sections given in table.\ref{tab:triHiggs_xsec_LHC} for various $\Lambda$'s that tri-Higgs production does not have much prospect at the $14$ TeV LHC. However, prospects of such process in a future $\sqrt{s}=100$ TeV proton-proton collider (FCC-hh) \cite{FCC:2018vvp,FCC:2018bvk,Benedikt:2020ejr} are really encouraging, particularly in scenarios like SM+H6 where cross-section is rather suppressed by the new physics scale. Even the first glimpse of triple Higgs events could be observed when expected upgradation of the LHC to a HE-LHC running at $27$ TeV becomes a reality. The cross-section numbers are displayed for HE-LHC for the center-of-mass energy $27$ TeV and FCC-hh for the center-of-mass energy $100$ TeV in table \ref{tab:trihiggs_xsec}. The tri-Higgs cross-section against the $hhh$ coupling at the LHC  $27$ TeV and $100$ TeV colliders are shown in Fig. \ref{fig:trihiggs}.
%%%%%%%%%
\begin{table}
\centering
    \begin{tabular}{|c|c|c|}
        \hline
        \textbf{$\Lambda$ (GeV)} & \textbf{$\kappa_{\lambda}$} & \textbf{$\sigma$ (fb)} \\
          \hline
        300 & -6.812 & 20.04 \\
        \hline 
        %350 & -4.74 & 8.82\\
        %\hline
        400 & -3.394  & 3.451\\
        \hline 
        500 & -1.81 & 1.104\\
        \hline 
        1000 & 0.29 & 0.1461\\
        \hline
       10000  & 0.992 &  0.0725\\
        \hline
    \end{tabular}
    \caption{Production cross-section of tri-Higgs via ggF ($gg \rightarrow hhh$) with renormalization scale set at $\frac{M_{hhh}}{2}$, where $M_{hhh}$ is the tri-Higgs invariant mass with $m_h=125 $ GeV at 14 TeV LHC for different values of the scale parameter $\Lambda$.}
    \label{tab:triHiggs_xsec_LHC}
\end{table}
%==============

%%%%%%%%%%%%%%%%%%
\begin{table}[h!]
    \centering
    
    \begin{tabular}{|c|c|c|}
        \hline
        \textbf{$\Lambda$ (GeV)} & \textbf{$k_{\lambda}$} & \textbf{$\sigma$ (fb)} \\
        
          \hline
        300 & -6.812 & 84.08 \\
        \hline
        350 & -4.739 & 31.64 \\
        \hline
        400 & -3.394  & 14.60 \\
        \hline
        450 & -2.472 & 7.71 \\
        \hline
        500 & -1.812 & 4.72 \\
        \hline
        550 & -1.324 & 3.51 \\
        \hline
         600 & -0.953 & 2.257 \\
         \hline
        650 & -0.664 & 1.72 \\
        \hline
        700 & -0.435 & 1.37  \\
        \hline
        750 & -0.250 & 1.14 \\
        \hline
        800 & -0.098 & 0.097 \\
        \hline

    \end{tabular}
    \caption{Production cross-sections of tri-Higgs via ggF ($gg \rightarrow hhh$) with renormalization scale $\frac{M_{hhh}}{2}$, where $M_{hhh}$ is the tri-Higgs invariant mass with $m_h=125 $ GeV at center-of-mass energy $27$ TeV at the HE-LHC for different values of the scale parameter $\Lambda$.}
    \label{tab:trihiggs_xsec}
\end{table}

%%%%%%%%%%%%%%$%%$

\begin{table}[h!]
    \centering
    \begin{tabular}{|c|c|c|}
     \hline
          \textbf{Collider} & \textbf{$\Lambda$ (GeV)} & \textbf{$\sigma$ (fb)} \\
        \hline     
		\multirow{3}{*}{HE-LHC} & 500 & 0.0214 \\ \cline{2-3} 
		& 600 & .0088 \\ \cline{2-3} 
		& 1000  & .0020 \\ \hline
    \hline     
		\multirow{3}{*}{FCC-hh} & 500 & 0.5029 \\ \cline{2-3} 
		& 600 & 0.2237 \\ \cline{2-3} 
		& 1000  & 0.0434  \\ \hline
    \end{tabular}

    \caption{Production cross-sections of four Higgses via ggF for various scale parameters $\Lambda$ ($gg \rightarrow hhhh$) with renormalization scale $\frac{M_{hhhh}}{2}$, where $M_{hhhh}$ is the four-higgs invariant mass with $m_h=125 $ GeV. The colliders namely HE-LHC and FCC-hh correspond to center-of-mass energies 27 TeV and 100 TeV respectively.}
    \label{tab:4-Higgs_xsec}
\end{table}
\subsection{Four-Higgs Production}\label{4-higgs}
\begin{figure}[h!]
	\centering
{\includegraphics[width=0.40\textwidth]{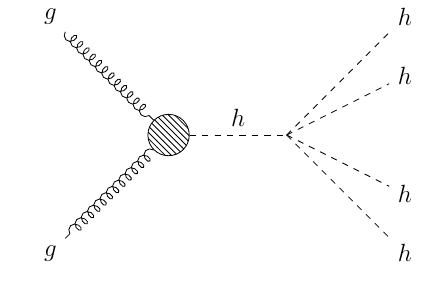}}
\caption{Typical Feynman diagram for four-Higgs production via ggF in SM+H6 scenario. The blob in the left vertex signifies toop and bottom quarks triangle loop. The right vertex is a typical vertex arising from SM+H6 model.} 
	\label{fig:multfey} 
\end{figure}

%\sout{The departure of SM+H6 model from the SM occurs in the structure of the scalar potential and thereby, in the appearance of new couplings in the Higgs sector. Particularly, the penta-Higgs and hexa-Higgs couplings appear as the concomitant effects of the presence of new physics beyond the scale $\Lambda$.}
The scalar potential of the SM+H6 scenario is remarkably different from the SM one leading to interesting Higgs phenomenology. Modified four-Higgs coupling and new five-Higgs coupling in the Higgs sector obtained in SM+H6 scenario may imprint new physics signatures. Particularly, five-Higgs coupling is absent in the SM because of renormalizability. An early, albeit small, excess in the production of four Higgs bosons at the collider may point towards such scenario, provided no new resonances show up at that energy. Contribution from new physics in four-Higgs production comes from the five-Higgs vertex where the single Higgs coming from ggF goes into four Higgses shown in Fig. \ref{fig:multfey}. Although, disentangling such contribution in the total four-Higgs cross-section is really challenging due to the presence of huge number of SM diagrams. Four-Higgs production cross-sections via gluon fusion ($gg \rightarrow hhhh$) at $27$ TeV HE-LHC and $100$ TeV FCC-hh are shown in table \ref{tab:4-Higgs_xsec}.

\section{Results from Gravitational Wave Astronomy}\label{sec-6}

% \textcolor{red}{The formulas below has been taken from $\textcolor{blue}{1811.09807}$ by vahid reza shajiee and ali tofighi.}

The thermal history of the EWPT can be analyzed once the tree level potential is set and the loop corrections, both at zero as well as non-zero temperatures, are performed to get the final form of the effective potential. The particular configuration of Higgs field $H$, called the critical bubble profile, obtained by solving the following equation of motion:
\begin{equation}\nonumber
\frac{d^2 H}{dr^2} + \frac{2}{r} \frac{dH}{dr} - \frac{\partial V_{\text{eff}}}{\partial r} = 0
\end{equation}
enters in the equation determining the three dimensional Euclidean action $S_3$ at finite temperature. The three dimensional Euclidean action is given by
\begin{equation} \label{eq:s3}
S_3(H,T) = \int~d^3x~\left[ \frac{1}{2} (\nabla H)^2 + (V_{\text{eff}}(H,T) - V_{\text{eff}}(0,T) ) \right].
\end{equation}
$V_{\text{eff}}$ used in both the above equations is taken from Eq.\ref{eq:veff_total}. 
This particular critical configuration, as the name suggests, minimizes the three dimensional Euclidean action $S_3$ which, in turn, determines the nucleation or tunneling rate from the formula $\Gamma(T) \sim \mathcal{A}(T)~e^{-S_3/T}$. In order to complete the EWPT throughout the entire Hubble volume, adequate bubble nucleation rate is required to win the race against the expansion rate of the universe. From the previous equation, it becomes obvious that the smallest value of the factor $\frac{S_3}{T}$ in the exponential term makes the largest contribution to tunneling. The numerical value which allows a SFOPT is 
$$
\frac{S_3}{T} \sim 140
$$
\cite{Apreda:2001us} and the temperature at which such $\mathcal{O}(1)$ nucleation rate is achieved is called the nucleation temperature $T_n$. 

During a first-order phase transition in the early universe, the transition does not occur uniformly across all space at once. Instead, regions of space undergo the transition in small localized patches, forming ``bubbles" of the new phase within the old phase. These bubbles then expand and eventually collide with each other and these collisions can generate GWs. The intensity and characteristics of these waves depend on factors like the energy difference between the phases and the dynamics of the bubble growth and collisions.

After the phase transition, the expanding bubbles induce sound waves (acoustic waves) in the surrounding plasma. These sound waves are essentially pressure waves moving through the hot, dense medium of the early universe. The coherent motion of the plasma due to sound waves can also generate GWs. The efficiency of sound wave production and their contribution to gravitational wave generation depends on how much of the energy of the phase transition is transferred into the kinetic motion of the plasma, as opposed to heating or other forms of energy dissipation.

In addition to bubble collisions and sound waves, the violent movement of the bubble wall through the plasma leads to turbulence in the plasma. Turbulence is a chaotic flow of the fluid (or plasma) that results from the nonlinear interactions of waves and shocks in the medium. In the presence of high magnetic fields, which are expected to be present in the relativistic charged plasma, this turbulence can become even more complex, leading to magnetohydrodynamic (MHD) turbulence which can efficiently generate GWs \cite{Caprini:2009yp}. In addition, we have assumed the thin-wall approximation \cite{Brown:2017cca,Megevand:2023nin} in which gives a faithful description of the phase transition process, provided the bubble wall achieves relativistic speed, which is often called the runaway scenario \cite{Megevand:2009gh} \footnote{Recent studies have shown that bubble wall reaches a terminal velocity due to the presence of considerable friction in the plasma \cite{Ellis:2019oqb, Ellis:2020nnr, Lewicki:2022pdb, Azatov:2024auq}. }. In this case, the total GW energy density comes from the three contributions: $\Omega_{GW} h^2 \simeq  \Omega_{turb} h^2 + \Omega_{sw} h^2 + \Omega_{col} h^2$. In this section $h$ stands for the Hubble constant taken to be $0.7$. We have used the fitted formula for the energy density carried by the sound wave:
%%%%%

\begin{eqnarray}
 \Omega_{sw} h^2 = 2.65 \times 10^{-6} \left(\frac{\beta}{H} \right)^{-1} v_b \left(\frac{ k_v \alpha}{1+\alpha} \right)^{2} \left(\frac{g_{*}}{100} \right)^{-1/3} \left(\frac{f}{f_{sw}}\right)^3 \left( \frac{7}{4 + 3 \left( \frac{f}{f_{sw}} \right)^2}  \right)^{7/2} 
\end{eqnarray}
Note that $H$ in the equation above and onwards, denotes the Hubble parameter, as opposed to Higgs field in Eq.\ref{eq:s3}.
Here, $f_{sw}$ is the peak frequency corresponding to the GW signal at present generated by the sound wave during the EWPT whose redhsifted formula is 
\begin{eqnarray}
 f_{sw}=1.9 \times 10^{-5}\frac{1}{v_b} \frac{\beta}{H} \frac{T_n}{100} \left(\frac{g_{*}}{100}\right)^{1/6}.
\end{eqnarray}
Similarly, the turbulence energy density modelled with a fitting formula is the following:
\begin{eqnarray}
 \Omega_{turb}h^2=3.35 \times 10^{-4} \left(\frac{\beta}{H} \right)^{-1} v_b \left(\frac{\epsilon k_v \alpha}{1+\alpha} \right)^{3/2} \left(\frac{g_{*}}{100} \right)^{-1/3} \frac{ \left( \frac{f}{f_{turb}} \right) ^{3}  \left( 1+ \frac{f}{f_{turb}} \right)^{-11/3}} { 1+ \frac{8 \pi f}{h_{*}} }   
\end{eqnarray}
with peak frequency $f_{turb}$ as the following
\begin{eqnarray}
  f_{turb}=2.5 \times 10^{-5} \frac{1}{v_b} \frac{\beta}{H} \frac{T_n}{100} \left(\frac{g_{*}}{100}\right)^{1/6}.
\end{eqnarray}
%%%
Finally the collision contribution in the energy density comes as
\begin{eqnarray}
 \Omega_{col} h^2 = 1.67 \times 10^{-5} \left(\frac{\beta}{H} \right)^{-2} \frac{0.11 v_b^3}{0.42 + v_b^2} \left(\frac{ k \alpha}{1+\alpha} \right)^{2} \left(\frac{g_{*}}{100} \right)^{-1/3} \frac{3.8 \left(\frac{f}{f_{col}} \right)^{2.8}}{1 + 2.8 \left(\frac{f}{f_{col}} \right)^{3.8}}   
\end{eqnarray}
where the corresponding peak frequency $f_{col}$ is
\begin{eqnarray}
 f_{col}=16.5 \times 10^{-6} \frac{0.62}{v_b^2-0.1v_b+1.8}\frac{\beta}{H} \frac{T_n}{100} \left(\frac{g_{*}}{100}\right)^{1/6}   
\end{eqnarray}

In this context, $g_{*}$ denotes the number of relativistic degrees of freedom present at the nucleation temperature $T_n$. For this analysis, we have set the parameter $\epsilon$ to a value of $0.1$.  Furthermore, the bubble wall velocity, denoted as $v_b$, is assumed to be approximately $1$, based on the scenario where the bubbles are considered to ``run away", meaning they expand rapidly with velocities approaching the speed of light. This assumption simplifies the dynamics of the phase transition, allowing us to focus on the key parameters of interest. $k$ and $k_v$ are parameters of the theory as given in \cite{Shajiee:2018jdq}.

%%%%%%%%5
\begin{figure}[t!]
    \centering
   % \subfloat[]	{\includegraphics[width=0.48\textwidth]{alpha_beta_H_6.pdf}}\label{fig:A2}
     \includegraphics[width=0.45\linewidth]{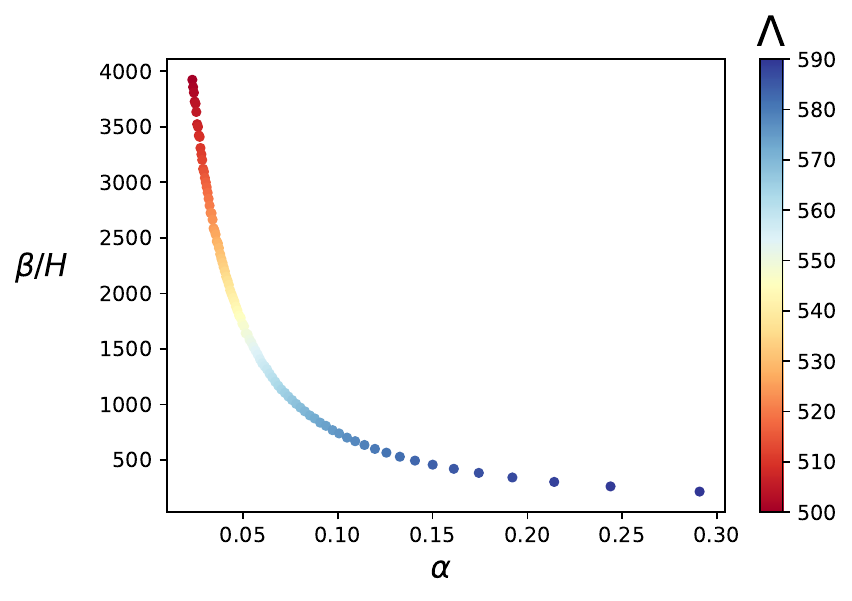}
     \includegraphics[width=0.45\linewidth]{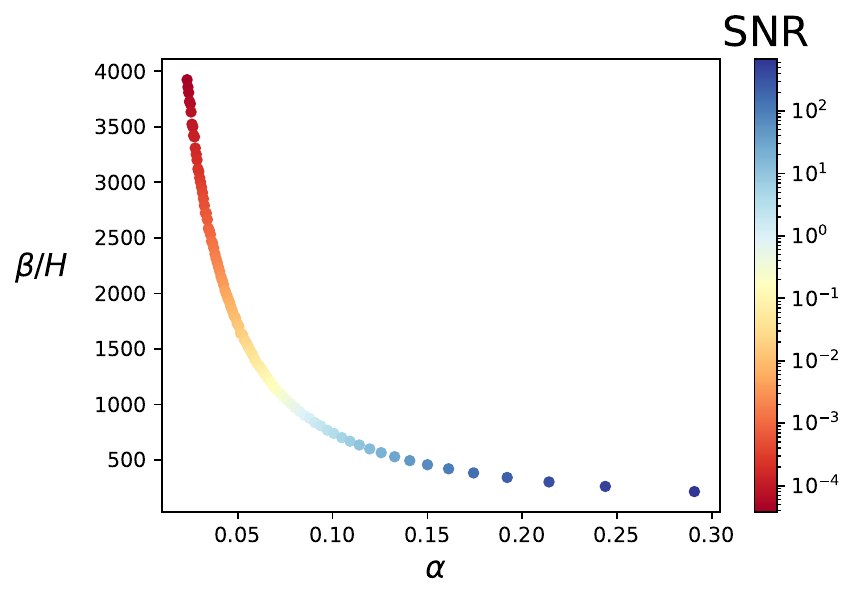}
    \caption{The two main parameters quantifying the GW signal are $\alpha$ and $\beta/H$, where $H$ is the Hubble parameter. Both the plots contain $\beta/H$ against $\alpha$ with the color gradient showing (i) the energy scale $\Lambda$ (in GeV), (ii) the SNR. Larger value of $\Lambda$ gives large SNR. In this case, $ 500 < \Lambda < 590$ is shown as it gives SFOPT within this range only. Here, $\Lambda$ is expressed in GeV.}
    \label{fig:GWalphabeta}
\end{figure}
%%%%%%
The strength of the GW signal produced in an early universe must be strong enough for detection in the present day, even after significant redshift. The detection prospect of GW can be determined by a parameter called signal-to-noise ratio (SNR) whose definition is \cite{Caprini:2015zlo,Schmitz:2020syl,Thrane:2013oya}
\begin{equation}
\text{SNR} = \sqrt{ \tau \int^{f_{max}}_{f_{min}} df~ \left[ \frac{\Omega_{\text{GW}(f)}}{\Omega_{\text{noise}(f)}}\right]^2}
\end{equation}
where $\Omega_{\text{noise}}$ is an experimental parameter. A signal may be caught in the detector if the SNR crosses a threshold value. This threshold, of course, a time dependent number with the advancement in technology and hence, the detector sensitivity, the number may come down to $\sim 10$. Note that the value of the SNR depends on a particular experiment. In this case, we have considered data from BBO \cite{Crowder:2005nr, Corbin:2005ny, Harry:2006fi}, DECIGO \cite{Seto:2001qf, Kawamura:2006up,Isoyama:2018rjb} and LISA \cite{Caprini:2015zlo, Athron:2023xlk} as the peak frequency in our scenario lies well within the sensitivities of these experiments.

\begin{figure}[htb]
    \centering
    \includegraphics[width=0.45\linewidth]{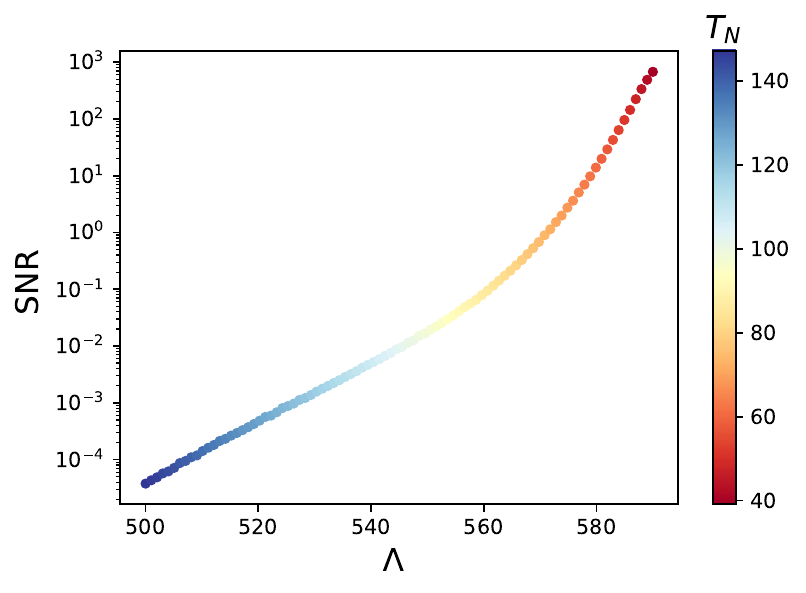}
    \includegraphics[width=0.45\linewidth]{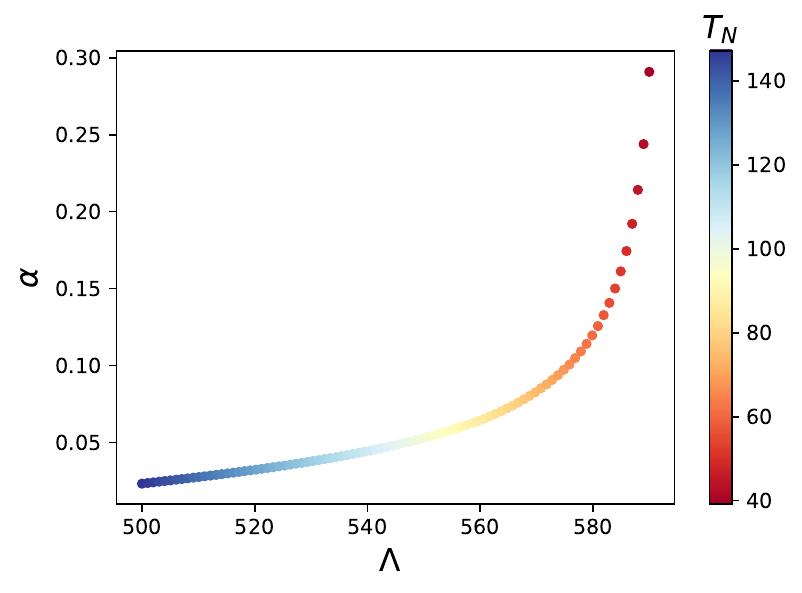}
    \caption{The variation of (i) SNR (left panel) (at LISA) and (ii) $\alpha$ (right panel) against $\Lambda$ (in GeV), the parameter of our theory, with the color gradient showing the nucleation temperature $T_N$. Large nucleation temperature is preferred by large $\Lambda$. The range is specified by the condition satisfying the SFOPT.}
    \label{fig:GWhr}
\end{figure}

The GWs can be produced by the collision of the bubbles at a temperature marginally lower than the nucleation temperature, often equated with the $T_n$. The (dimensionless) parameters quantifying the quality of the GW spectra are:
\begin{enumerate}
    \item The strength of the transition defined as 
$$
\alpha = \frac{\Delta \rho}{\rho_R} = \left. \frac{1}{\rho_R} \left[ \Delta V - T \Delta \frac{dV}{dT} \right] \right\vert_{T=T_n}
$$
where $\Delta \rho$ signifies the difference in energy density between the true and false vacuum which is released after the transition and $\rho_R$ is the radiation energy density. $\Delta V$ is the difference between minima of the scalar potential. 

    \item The nucleation rate or, the span of transition given by
$$
\beta = \left. H_n T_n \frac{d(S_3/T)}{dT}\right\vert_{T=T_n}
$$
\end{enumerate} 
A large $\alpha$ and a smaller $\beta$ would typically results in good SNR. From Fig.\ref{fig:GWalphabeta}, we can see that the larger value of the scale parameter $\Lambda$ provides good SNR. However, that a smaller nucleation temperature is preferred by larger $\Lambda$ can be observed from Fig.\ref{fig:GWhr}. Therefore, from the Figs. \ref{fig:GWalphabeta} and \ref{fig:GWhr}, it can be safely stated that new physics scale on the larger side would give large $\alpha$ and small $\beta/H$ which, in turn, are required for a decent SNR signifying a detectable GW signal at the LISA \cite{Caprini:2015zlo, Athron:2023xlk}, BBO \cite{Crowder:2005nr, Corbin:2005ny, Harry:2006fi} and DECIGO \cite{Seto:2001qf, Kawamura:2006up,Isoyama:2018rjb}.

%%%%%%%%%%%
%%%%%%%%%%

To understand how this energy is distributed across different frequencies, we define the energy density spectrum $\Omega_{GW}(f)$. This is a dimensionless quantity that represents the fraction of the total energy density in the universe contributed by GW per logarithmic frequency interval. It is given by:
$$
\Omega_{GW}(f) = \frac{1}{\rho_c}~\frac{d\rho_{GW}}{d \text{ln} f}.
$$
Here, 
\begin{itemize}
    \item $f$ is the frequency of the GWs.
    \item $\rho_c$ is the critical energy density of the universe (the energy density required for a flat universe)
    \item $ \frac{d\rho_{GW}}{d \text{ln} f}$ is the energy density of GWs per logarithmic frequency interval
\end{itemize}
%%%
\begin{figure}[h!]
    \centering
    \includegraphics[width=0.6\linewidth]{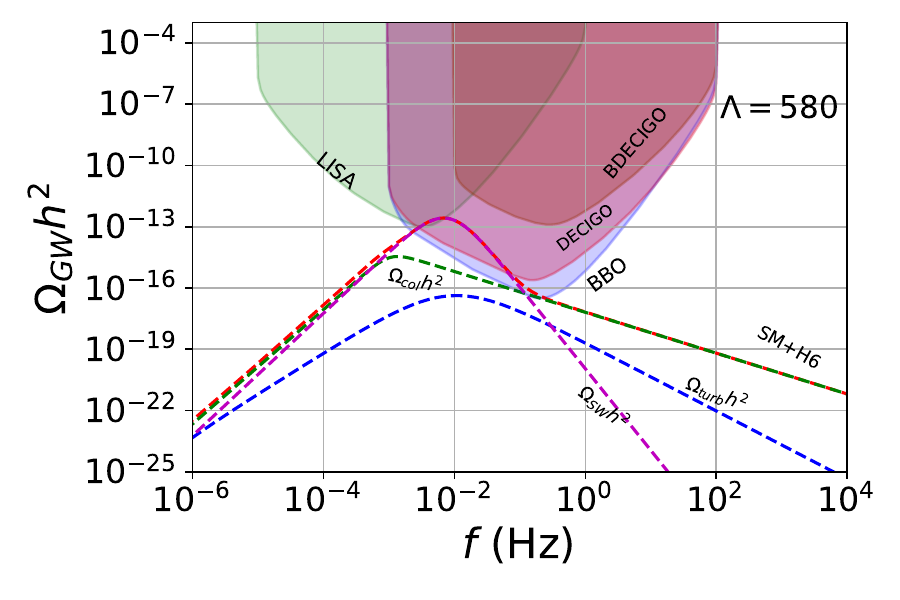}
    \caption{The GW signal as a function of frequency for $\Lambda = 580$ GeV for SM+H6 scenario. The sound wave, collision and turbulence contributions adding to the total energy density spectrum are also shown. The nucleation temperature is around $60$ GeV. It can be seen that the peak frequency is well under the LISA \cite{Caprini:2015zlo, Athron:2023xlk}, BBO \cite{Crowder:2005nr, Corbin:2005ny, Harry:2006fi} and DECIGO \cite{Seto:2001qf, Kawamura:2006up,Isoyama:2018rjb} detector sensitivities.}
    \label{fig:GWden}
\end{figure}
%%%%%
Gravitational waves generated during early universe processes, such as phase transition, typically produce a broad spectrum that can extend over a wide range of frequency. The exact shape of $\Omega_{GW}(f)$ depends on the detailed dynamics of various processes, including bubble collisions, sound waves, and turbulence from phase transitions. For many sources of GWs, the energy density spectrum can be modeled as a power-law function of frequency: $\Omega_{GW}(f) \propto f^n$, $n$ being the power-law index. These typically produce a broad spectrum with a peak frequency determined by the temperature and dynamics of the phase transition. The energy density may increase at low frequencies, reach a peak, and then decay at higher frequencies, as can be seen in Fig.\ref{fig:GWden}. GW energy density spectrum against frequency for model parameter $\Lambda = 580$ GeV is displayed in Fig.\ref{fig:GWden}. The individual contributions for turbulence, bubble collision and sound waves are also shown the same plot. The sound wave contribution can be seen to be the dominant one. The peak frequency of the cumulative contribution is well within the reach of LISA \cite{Caprini:2015zlo, Athron:2023xlk}, BBO \cite{Crowder:2005nr, Corbin:2005ny, Harry:2006fi} and DECIGO \cite{Seto:2001qf, Kawamura:2006up,Isoyama:2018rjb}. In Fig.\ref{fig:GWdenmult}, we have displayed the shift in the peak frequency of GW spectra for different benchmark values of the scale parameter $\Lambda$ for SM+H6 scenario. We observe larger $\Lambda$ favours lower nucleation temperature ($T_n$) (and equivalently, smaller peak frequency) which can also be corroborated by the Fig.\ref{fig:GWhr}. The nucleation temperatures $T_n = 137.2, 97.3, 60.2, 39.1 $ (all in GeV) used in Fig.\ref{fig:GWdenmult} correspond to $\Lambda = 510, 550, 580, 590$ (all in GeV).

%%%%%%%%

\section{Summary and Conclusions}\label{sec-7}
%%%
%%%%
\begin{figure}[h!]
    \centering
    \includegraphics[width=0.6\linewidth]{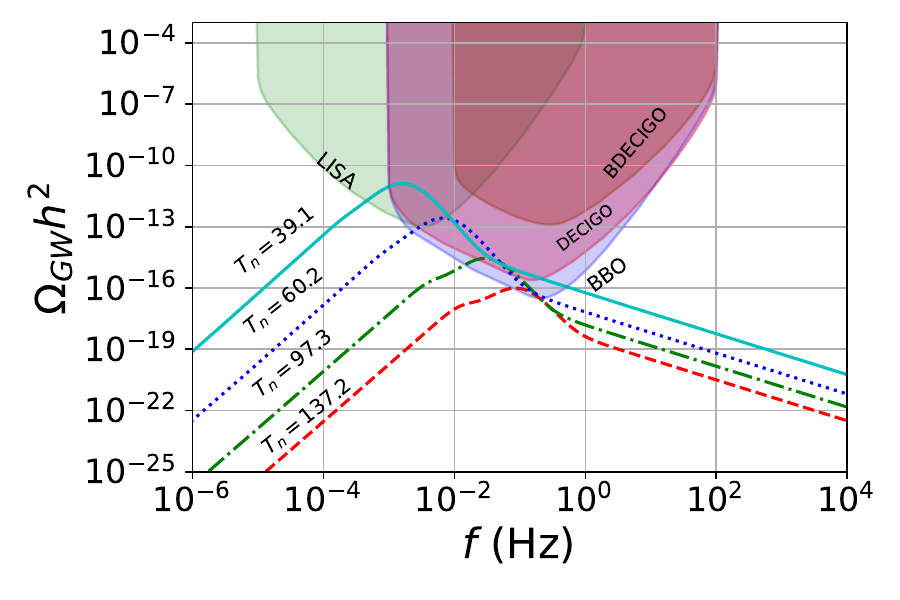}
    \caption{The GW spectra as a function of frequency for some benchmark values of the scale parameter $\Lambda$ for SM+H6 scenario. The curves corresponding to lower nucleation temperatures ($T_n$) or, equivalently lower peak frequencies correspond to higher values of $\Lambda$. The integrated detector sensitivities of LISA \cite{Caprini:2015zlo, Athron:2023xlk}, BBO \cite{Crowder:2005nr, Corbin:2005ny, Harry:2006fi} and DECIGO \cite{Seto:2001qf, Kawamura:2006up,Isoyama:2018rjb} are also shown. All $T_n$s in the figure are in GeV.}
    \label{fig:GWdenmult}
\end{figure}

%%%%%
The absence of new resonances has naturally pushed us into thinking in terms of effective theory. The inclusion of a dimension six operator $((H^\dagger H)^3)$ in the Lagrangian offers some insight of the unknown new physics beyond some energy scale $\Lambda$ at the expense of renormalizability of the theory. Such term in the Higgs potential provides a concrete scenario where modifications to the Higgs sector can lead to observable consequences in both gravitational wave astronomy and collider experiments. By studying the impact of this operator on the electroweak phase transition and exploring multi-Higgs signatures at colliders, we can gain deeper insights into the nature of the Higgs field and the potential existence of new physics beyond the Standard Model. 
%With no new particles coming up at the LHC, the primary focus of the experimental particle physics has bifurcated into exploring the sky for garnering astrophysical and cosmological data, and various low energy experiments. Furthermore, absence of new resonances has naturally pushed us into thinking in terms of effective theory. The inclusion of dimension six term in the Lagrangian offers some insight of the unknown new physics beyond some energy scale $\Lambda$ at the expense of renormalizability of the theory.\\
\noindent
In SM+H6 scenario the triple Higgs and quartic Higgs couplings get modified with respect to the SM. Triple Higgs coupling is already under probe through the di-Higgs searches at the LHC. Moreover, SM+H6 model also predicts five-Higgs and six-Higgs couplings which are absent in the SM.\\
In Fig.\ref{fig:lhc_dihiggs} the left panel showcases the the latest bound on self-coupling modifier $\kappa_{\lambda}$, defined as $\kappa_{\lambda} =\frac{ \lambda^{SM+H6}_{hhh}}{\lambda^{SM}_{hhh}}$ from di-Higgs production at 13 TeV LHC Run II performed by the ATLAS collaboration in SM+H6 model. The constraint on the coupling modifier $\kappa_{\lambda}$ is allowed in the range $ -5.09 < \kappa_{\lambda} < 0.998$, where the upper limit basically corresponds to the SM. The right panel shows the same bound in terms of the scale parameter $\Lambda$ for the SM+H6 model. Equivalently, we translate di-Higgs cross-section bound in terms of $\Lambda$,  to be  $ \Lambda \gtrsim 340 ~\text{GeV}$. The upper limit of $\Lambda \approx 20$ TeV where the scenario reduces to SM.\\
We have also presented the projected cross-sections for the di-Higgs production from gluon-gluon fusion (ggF) and vector boson fusion (VBF) at 14 TeV LHC and 27 TeV HE-LHC in Fig.\ref{fig:lhc_plot2}.
 In addition Fig.\ref{fig:multfey} displays the production cross-sections for tri-Higgs from ggF ($gg\rightarrow hhh$) at 27 TeV HE-LHC and 100 TeV FCC-hh which are still very small due to destructive interference of various Feynman diagrams. In table \ref{tab:4-Higgs_xsec} we have given the production cross-section for four-Higgs via ggF ($gg \rightarrow hhhh$) at $27$ TeV HE-LHC and $100$ TeV FCC-hh for some representative values of the scale parameter $\Lambda$.\\
\noindent
\noindent
We have followed the most well-known formalism of estimating the effective potential by means of loop corrected potentials at zero and finite temperature.
On top of that, the requirement of SFOPT to generate detectable stochastic GW from the collision of bubbles of broken EW phase, restricts the scale parameter within $450~\text{GeV} \lesssim \Lambda \lesssim 590$~GeV. 

\noindent
We have found the numerical values of $\alpha$ and $\beta/H$ for various $\Lambda$ and also computed what is called a good signal in terms of a quantity called SNR (see Fig. \ref{fig:GWalphabeta} and \ref{fig:GWhr} respectively). The peak frequency of resulting GW signal including all three types of contributions can be found (see Fig. \ref{fig:GWden}) to fall within sensitivity reach of LISA  \cite{Caprini:2015zlo, Athron:2023xlk}, BBO \cite{Crowder:2005nr, Corbin:2005ny, Harry:2006fi} and DECIGO \cite{Seto:2001qf, Kawamura:2006up,Isoyama:2018rjb} experiments. Higher new physics scale favours lower nucleation temperature, and hence, satisfies the condition for strongly supercooled state in the early universe, which is required for stronger GW spectrum (see Fig.\ref{fig:GWdenmult}).

\noindent
In the absence of new physics beyond the SM, the presence of any new physics can be made condensed in the higher-dimensional operator whose effects is hard to observe at the LHC, due to its scale suppression. Therefore, an early result of such scenarios could be expected from GW astronomy searches which are already undergoing or soon to be coming, before any direct measurement can be done in future colliders. 

\vspace{1cm}
\section*{Acknowledgements}

Authors thank Md. Raju for fruitful discussions on NLOCT package. AM and SN are supported by the Science and Engineering Research Board (SERB) (Anusandhan National Research Foundation), Govt. of India grant number SUR/2022/001404.

\bibliographystyle{hephys}
\bibliography{sample}

\end{document}